# Scratching lithography, manipulation, and soldering of 2D materials using microneedle probes

Qing Rao[1,2,#], Guoyun Gao[1,3,#], Xinyu Wang[1,2], Hongxia Xue[1,2,*], and Dong-Keun Ki[1,2,*]

[1]Department of Physics, The University of Hong Kong, Pokfulam Road, Hong Kong, China

[2]HK Institute of Quantum Science & Technology, The University of Hong Kong, Pokfulam Road, Hong Kong, China

[3]Department of Electrical and Electronic Engineering, The University of Hong Kong, Pokfulam Road, Hong Kong, China

[#]Equal Contributions

[*]Correspondence Authors

Email: hxxue@hku.hk; dkki@hku.hk

## Abstract

We demonstrate a facile technique to scratch, manipulate, and solder exfoliated flakes of layered 2D materials using a microneedle probe attached to the precision *xyz* manipulators under an optical microscope. We show that the probe can be used to scratch the flakes into a designated shape with a precision at micrometer scales, move, rotate, roll-up, and exfoliate the flakes to help building various types of heterostructures, and form electric contacts by directly drawing/placing thin metal wires over the flake. All these can be done without lithography and etching steps that often take long processing time and involve harmful chemicals. Moreover, the setup can be easily integrated into any van der Waals assembly systems such as those in a glove box for handling air/chemical-sensitive materials. The microneedle technique demonstrated in this study therefore enables quick fabrications of devices from diverse 2D materials for testing their properties at an early stage of research before conducting more advanced studies and helps to build different types of van der Waals heterostructures.

Successful exfoliation of graphene from graphite in 2004 has boosted extensive efforts to search for new layered 2D materials with diverse electronic properties[1]. It has led to identifications of over 1,000 2D materials in theory[2-7] but with limited experimental verifications. Different techniques have been developed to help flake exfoliation[8-14], vdW assembly[15-21], and device fabrication[10,11,22-28], but they often need advanced fabrication process or equipment, such as fabrication of the stencil mask[10,11,24,28], metal deposition in a glove box[10,11,28], AFM[29-42], or a high-power laser integrated in an optical microscope[43]. It is therefore beneficial to establish a facile and cost-effective technique to quickly fabricate samples from different 2D materials for initial testing or to build vdW heterostructures.

Here, we demonstrate that a commercially available microneedle probe (originally manufactured for probe stations) can be used to scratch 2D materials into a designed shape (**Figs. 1** and **2**), exfoliate, roll-up, move, and rotate the flakes to create or modify different types of heterostructures (**Figs. 3** and **4**). We can also use the probe to draw or solder thin metallic wires directly on the exfoliated flakes to form electrical contacts (**Fig. 5**). The successfully demonstrated tasks and comparison with other methods[20,21,29-45] are summarized in **Table I**. Essentially, our microneedle probe system is identical to the vdW transfer system without the need for adding new sophisticated equipment like a high-power laser to the systems[43]. The setup can therefore be conveniently integrated in a glove box for handling air/chemical-sensitive materials[46-48] (see **Fig. 1(a); Fig. S1**).

**Table I.** Comparison between the microneedle technique and other manipulation methods.

| Technique | Scratching/ Pattering | Rolling | Exfoliating/ Thinning | Rotating /Moving | Soldering | Working area |
|---|---|---|---|---|---|---|
| Microneedle | Possible[45] | Possible | Possible | Possible | Possible[44] | ~ 100 mm$^2$ |
| Microdome[20,21] | Not possible | Possible | possible | Possible | Not possible | ~ 100 mm$^2$ |
| AFM[29-42] | Precise | Not shown | Possible | Precise | Not possible | ~ 0.1 mm$^2$ |
| Laser[43] | Possible | Not possible | Not possible | Not possible | Not possible | ~ 100 mm$^2$ |

Among various functions of the microneedle probe system, we begin with the scratching lithography, a direct patterning method to define the shapes of the thin flakes[45] as shown in **Fig. 1**. In the setup, a mechanically strong tungsten microneedle probe with a tip diameter of 50~200 nm is attached to the *xyz* manipulator stage of the vdW transfer system. This allows us not only to define the shapes with micron precision but also to control the force to scratch the flake depending on the material and thickness. After mounting a silicon substrate with the flake for scratching on the sample stage, we carefully approach the probe tip to the flake. Once the tip touches the flake, we move the tip in *xy* direction to scratch the flake along the designed paths (see **supplementary material** for more details).

**Fig. 1(b)** shows a Hall-bar patterned monolayer graphene by scratching lithography. We were able to realize Hall probes with widths and distances down to around 1 µm, which is enough for most of the 2D materials research. Notably, similar to the graphene flakes cut by AFM tips[31,39], we found that the scratched edges are smoother when the flake is scratched along the crystalline direction as indicated by black arrows and dashed lines in **Fig. 1(b)** than when scratched along an arbitrary direction indicated by white arrows. This can be attributed to a smaller critical stress needed to crack graphene along armchair or zigzag edge[49,50]: the graphene edges torn by the microneedle probe tend to align with the energetically preferred armchair or zigzag directions, leading to a smoother edge when the needle scratches along the same direction.

When integrated in a glove box (see **Fig. S1(b)**), the scratching lithography setup can help shaping air/chemical-sensitive materials, such as black phosphorus, 1T'-WTe$_2$, 2H-MoTe$_2$, and most of 2D magnets[46,47,51-53], into a desired structure. In this way, one can protect both the basal plane and the edges of the flake from harmful environments during fabrication and measurement while the device geometry is defined. **Fig. 1(c)** shows few layers of 2H-MoTe$_2$ that are directly patterned into a rectangle for making a van der Pauw geometry[54,55] (top) or into a Hall bar (bottom) by the scratching lithography before being sandwiched by the two hBN flakes inside a glove box.

The technique can also be applied for cut-and-stack of 2D materials[37,38,43] to precisely control twisting angle when studying emerging twistronics[56]. The inset of **Fig. 1(d)** shows the image of a bilayer graphene flake scratched in the middle by the microneedle probe before being assembled into a twisted double bilayer graphene device. The back gate voltage ($V_{bg}$) dependence of the four-terminal resistance ($R_{xx}$) of the device (**Fig. 1(d)**; **see supplementary material**) shows the two characteristic satellite resistance peaks (marked by black triangles) resulting from the formation of secondary Dirac cones in the moiré pattern, and the twist angle is estimated to be around 0.45° which is close to the targeted value 0.50°. Compared with the AFM[37,38] or laser cutting[43] methods which require advanced equipment, the scratching lithography provides an accessible and rapid approach for shaping the flakes.

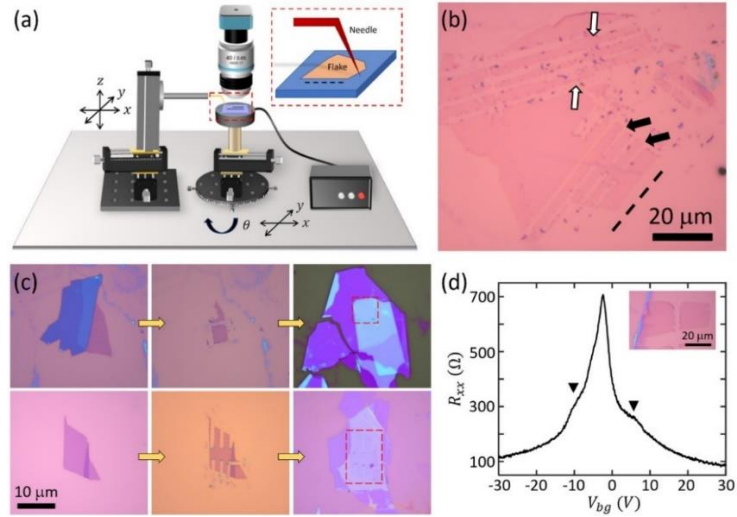

**Figure 1.** Scratching lithography. (a) Schematic of the microneedle probe system, which is the vdW transfer system with a microneedle probe holder replacing the glass slide holder. The inset shows a zoomed-in view of the schematic. (b) The optical microscope image of a graphene flake patterned into a hall bar by scratching. The black dashed line denotes the crystalline direction of the flake whereas the black and white arrows indicate the edges scratched along the crystalline and arbitrary directions, respectively. (c) Illustration of shaping air/chemical-sensitive 2H-MoTe₂ flakes into a rectangle and a Hall bar inside a glovebox (top and bottom respectively) before hBN encapsulation. (d) The back gate voltage ($V_{bg}$) dependence of four-terminal resistance ($R_{xx}$) of a twisted double bilayer graphene device made by the cut-and-stack method using the microneedle probe. The two black down triangles mark the satellite peaks from which we estimate a twist angle of 0.45°. Inset: an optical microscope image of the bilayer graphene cut by the microneedle probe.

Thanks to the simplicity of the setup, one can also integrate an electrical measurement unit to monitor the change of the resistance of the flake while scratching. **Fig. 2(a)** shows a schematic of the combined system, where we replace the normal sample stage with a chip carrier holder, which is electrically connected to measurement instruments through a breakout box (not shown). We demonstrate the technique using graphene device (**Fig. 2(b)**). After fabricating electrical contacts (5 nm Cr/50 nm Au) on graphene exfoliated on a silicon substrate using conventional lithography and lift-off processes, we placed the device on the chip carrier holder and measure its resistance $R_{ab,fe} \equiv V_{fe}/I_{ab}$ by applying a low-frequency AC current from the contact $a$ to $b$ and measuring the voltage between the contact $f$ and $e$ (**Fig. 2(b)**). Meanwhile, we use the microneedle to scratch the graphene flake between the contacts $f$ and $e$. As shown in **Fig. 2(c)**, during the whole scratching process, we were able to trace the change of the resistance in time precisely until the flake is cut completely (the inset of **Fig. 2(c)**). Since the conductance value is proportional to channel width, from the ratio of conductance at step 1 and 8 (before and after scratching), we can also roughly estimate that the smallest achievable constriction is less than 40 nm. Moreover, for another pair of contacts ($d$ and $e$), we paused scratching at each step and measured $V_{bg}$ dependence of the conductance $G_{bc,ed} \equiv I_{bc}/V_{ed}$ to show a stronger suppression of the conductance in a wider $V_{bg}$ range (**Fig. 2(d)**). This is consistent with the wider gap opening for the narrower graphene[57,58].

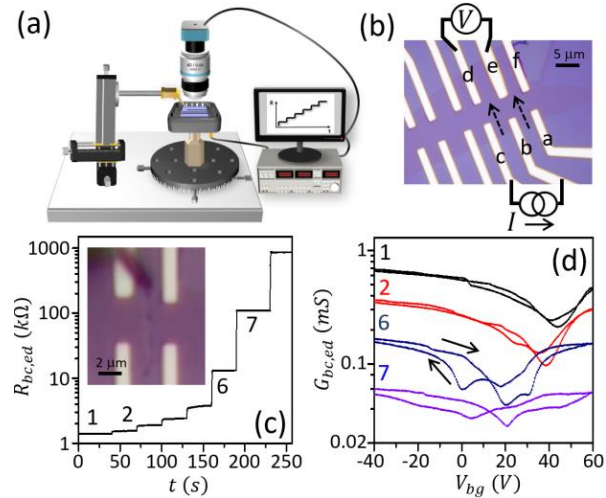

**Figure 2.** In-situ electrical measurement. (a) Schematic of an in-situ electrical measurement system that allows one to measure electrical signal from the flakes while scratching. (b) An optical microscope image of the graphene device with a four-terminal measurement configuration used in (c) and (d). The dashed arrows indicate the position and direction of scratching. (c) The evolution of the resistance $R_{ab,fe} \equiv V_{fe}/I_{ab}$ in time while cutting the flake between the contacts $e$ and $f$ (the inset: the device image during scratching). The number indicates a scratching step. (d) The $V_{bg}$ dependence of the conductance $G_{bc,ed} \equiv 1/R_{bc,ed}$ measured after each step while scratching the flake between contacts $e$ and $d$.

This in-situ electrical measurement technique can be used to fabricate well-defined narrow constrictions in 2D materials, such as narrow Hall probes[59], quantum point contacts[39,42,60], or quantum dots[61,62]. It also offers an interesting direction to study 2D semiconductors. For example, while measuring field-effect-transistor (FET) characteristics of the 2D semiconductor, one can scratch the flake in steps to study the channel-width dependence of the FET characteristics. We therefore believe it can offer new means to fabricate 2D materials devices and probe their electrical properties simultaneously.

Unlike the AFM tip[32,33,40] or laser, the microneedle probe can also exfoliate or roll up thick flakes as large as few tens of micrometers quickly because its tip is harder and heavier than the AFM tip such that it can easily break the vdW interaction between the layers. **Fig. 3(a)** show that few layers of thick graphite flakes can be peeled off from the top by the microneedle probe without bottom layers damaged. For this, we lowered the tip towards the flake slowly while scanning across the flake in a lateral direction. When the tip touches the top of the flake, we could see some layers begin to be peeled off. The tip can be lowered further if we want to thin down the flake more. Currently, the technique is limited by the sharpness of the tip (50~200 nm in diameter), so the precise control over thickness is not easy, and we could exfoliate only thick flakes with thicknesses over few tens of nanometers. This can be further improved by using a much sharper needle or by tilting the tip to make its end touch the flake boundary. Nevertheless, the entire process is quick and does not require sophisticated equipment like AFM. Moreover, it can be used to reduce the thicknesses of non-vdW 2D crystals such as $ZrN$[63,64] or $BaSnO3$ (LBSO)[65,66], whose electric or optical properties depend on their thicknesses in the range of 50 to hundreds of nanometers, but are difficult to be exfoliated using conventional mechanical exfoliation methods or by AFM.

By lowering the tip further down, we can also roll-up the flakes from the silicon substrates completely as shown in **Fig. 3(b)**, where the thick graphite flake was rolled up and pushed away from its original location. This can be used to make a scroll of 2D material flake or to clean up unwanted thick flakes around the thin flake of interest (see **Fig. 3(c)**). The latter is particularly useful in vdW assembly as mechanical exfoliation often produces thin layers of 2D materials attached to or surrounded by thick flakes, which hinders an adhesion of the thin flake to the polymer or the pick-up flake during the vdW transfer.

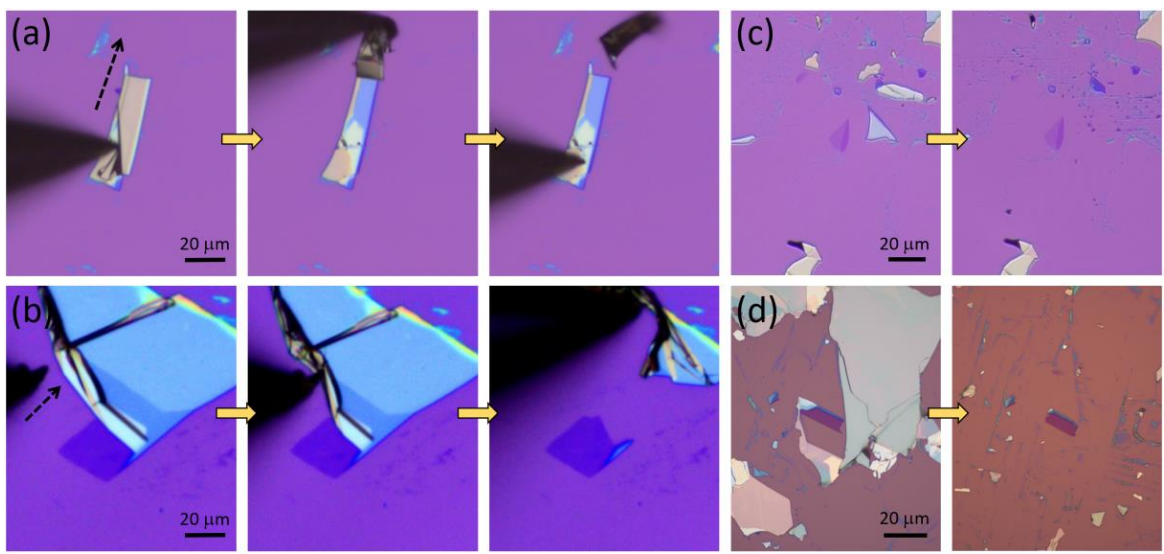

**Figure 3.** Exfoliation and roll-up. (a) The exfoliation of a thick graphite flake. The top layers are peeled off by the microneedle probe, while the bottom layers remain unaffected. (b) The roll-up and removal of a thick graphite flake attached to a few-layer graphene flake. (c,d) The optical microscope images of few layer graphene and $MoS_2$, respectively, before and after the surrounding thick flakes are cleaned up by the microneedle probe.

Interestingly, we found that when the flakes are placed on atomically flat surfaces like graphite or hBN, the microneedle often rotates or slides the flakes rather than scratching or peeling them off (see **Fig. 4**). This can be understood by the fact that on silicon substrate, the adhesion between the flakes and the amorphous silicon oxide layer is not uniform so when the needle touches the flake, some parts of the flakes are fixed strongly, leading to scratching or peeling off (see **Fig. 4(c)**). On the other hand, on an atomically smooth surface, the adhesion between the flake and the substrate becomes uniform such that the entire flake can be moved together with the needle. **Figs. 4(b)** further show that the microneedle can also rotate and slide the pre-patterned flake, similar to the AFM tip[35,36]. The technique works as long as the probe does not cross the edge of the atomically flat substrate or touch its surface (i.e., the graphite in **Figs. 4(a-b)**). This is consistent with our finding that the hBN flakes that can be rotated or moved are generally thicker than ~20-30 nm which is close to the sharpness of the tip.

We note that compared with the AFM tip[35,36] or PDMS hemisphere[20,21], the microneedle can rotate or slide the flake in a much longer distance, limited only by the area of the atomically smooth surface. The capability to rotate or slide the flakes in such a long distance opens new opportunities for vdW engineering. For example, one can turn on and off Andreev reflection by moving superconductor flake(s) onto or away from the metallic bottom layers[67] or place 2D magnets at locations where one wants to induce or detect magnetism, such as at the middle or the edge of graphene Hall bars[68]. Further, as demonstrated in **Fig. 2**, we can also integrate the in-situ electrical measurement setup to study electrical properties of the vdW heterostructures during manipulation.

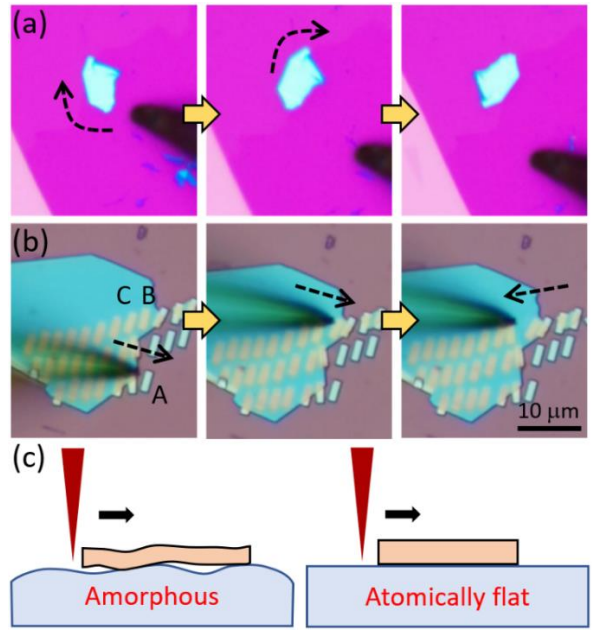

**Figure 4.** Rotation and sliding. (a,b) Rotation and sliding of as-exfoliated and pre-patterned hBN flakes on a graphite surface by a microneedle probe, respectively. The dashed arrows in (b) mark the direction of the tip movement. Only those on the atomically flat graphite surface (the flakes B and C) can be moved. (c) Comparison of the interface between the flake and the substrate with amorphous and atomically flat surface, respectively from left to right.

Lastly, we demonstrate a direct soldering of metal wires to the flakes using a microneedle probe. For this, we first use Field's metal (alloys of indium and tin) as it has low melting temperature (~50 °C) that is beneficial for 2D materials research in many aspects like in keeping twist angle in moiré structures[69]. The detailed operation is illustrated in **Fig. 5(a)**, where we place a small piece of Field's metal on the substrate away from the target flake, heat up the substrate to 60 °C above the metal's melting point, and draw a wire from the melted droplet towards the target by the microneedle to form contacts (inset of **Fig. 5(b)**). Alternatively, we can also follow the soldering method reported previously[44] by pulling out spikes from melted indium beads using a microneedle, bringing the spike to the flake, and raising the needle to release the spike on the flake (inset of **Fig. 5(c)**). Interestingly, we found it much more difficult to pull out the spikes from Field's metal than indium. It can be due to the comparable heat capacity[70] but lower melting point of Field's metal such that it gets harder to be solidified when pulled out from the melted droplet.

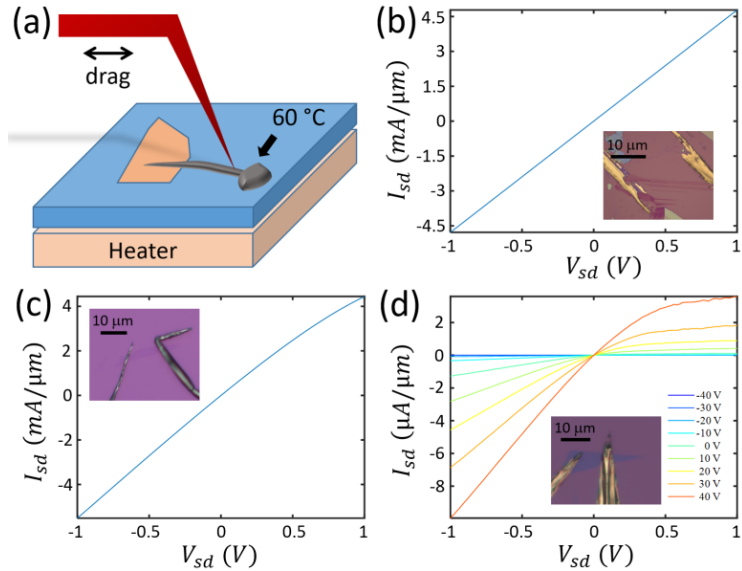

**Figure 5.** Soldering. (a) Schematic of soldering. Metal wires are drawn from the melted droplets to the flakes on silicon substrate by a microneedle probe to form contacts. (b,c) The characteristic $I-V$ curves (at zero gate) of two graphene devices soldered by Field's metal and by indium spikes, respectively. The image of the corresponding device is shown in the insets. (d) The image and characteristic $I-V$ curves of a soldered MoS$_2$ device by indium spikes as a function of back gate voltage.

To demonstrate electrical contacts between the microneedle-soldered wires and 2D materials, we soldered Field's metal and indium wires to graphene monolayers and measured their current-voltage ($I-V$) characteristics at zero back gate. The results are shown in **Figs. 5(b,c)** for Field's metal and indium, respectively, displaying nearly linear $I-V$ curves in both cases from the formation of Ohmic contacts. Additionally, **Fig. 5(d)** shows the result from monolayer MoS$_2$ soldered by indium wires, which exhibit a n-type FET behavior as found in exfoliated flakes[71], with an on-off ratio exceeding $10^6$ and a large on-current density of ~10 µA/µm at 40 V back gate and source-drain voltage $V_{sd} = -1$ V. The current saturation at positive $V_{sd}$ indicates the possible formation of Schottky barrier at the indium and MoS$_2$ interface which needs to be investigated further.

In summary, we have demonstrated a microneedle probe system with multiple functionalities in fabrication and electrical characterizations of devices based on 2D materials. We show that using a microneedle probe, one can 1) pattern thin flakes (<50 nm for graphite) by scratching (**Fig. 1**), 2) remove or thin down thick flakes by rolling or exfoliation (**Fig. 3**), 3) move or rotate the topmost layers in vdW heterostructures (**Fig. 4**), and 4) form electrical contacts by drawing thin metal wires (**Fig. 5**). All these operations can be conducted in existing vdW transfer setups by simply adding a microneedle and its holder to the manipulator stage, and can be done in a glove box with controlled environments for handling air/chemical-sensitive materials. One can also add a measurement unit for in-situ electrical characterizations during the manipulation (**Fig. 2**). Compared with other known lithography-free fabrication[10,11,23-28] and manipulation methods[20,21,29-45], the microneedle technique has its own advantages in flexibility, cost, and efficiency, while offering a large working area with sufficient precision (see **Table I**). We believe the versatility and simplicity of the microneedle technique can lead to more accessible and rapid device fabrication, as well as provide new strategies in vdW engineering.

**Acknowledgements**

The work is financially supported by the National Key R&D Program of China (2020YFA0309600) and by the University Grants Committee/Research Grant Council of Hong Kong SAR under schemes of Area of Excellence (AoE/P-701/20), CRF (C7037-22G), ECS (27300819), and GRF (17300020, 17300521, and 17309722). We also acknowledge financial support from University Research Committee (URC) of The University of Hong Kong under the schemes of Seed Fund for Basic Research (202111159043) and Seed Funding for Strategic Interdisciplinary Research Scheme.

**Conflict of Interest**

The authors have no conflicts to disclose.

**Data Availability**

The data that support the findings of this study are openly available in the HKU data repository at https://doi.org/10.25442/hku.24005367.v1.

# Supplementary Material

**Microneedle probe system**

The microneedle probe system used in this study consists of an optical microscope, a sample stage with $xy\theta$ manipulators and a PID-controlled heater, and another $xyz$ manipulators with a home-made microneedle probe holder to fix the microneedle probe (see **Fig. 1(a)**). The real pictures of the two setups placed in air and glove box are shown in **Fig. S1(a)** and **(b)**, respectively.

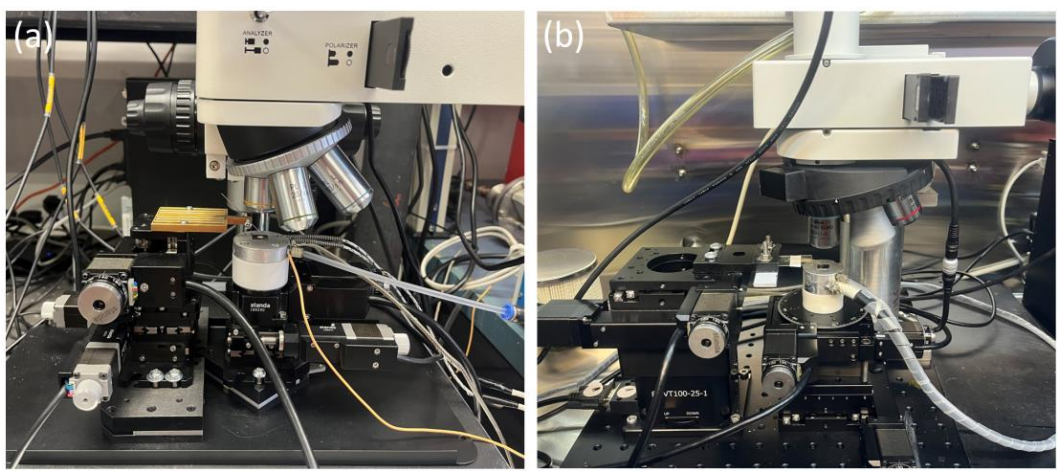

**Figure S1.** Photos of the microneedle probe systems in (a) air and (b) glove box.

To scratch/exfoliate/move/rotate the flake, we first place the tip of the microneedle probe directly over or near the flake depending on the purpose, and lower it down slowly using the $z$-translation stage until it touches the substrate or the flake surface. Once the tip touches the surface which can be noticed by checking if the tip is horizontally moving or not, we adjust the $xy$-translation stage to move it to scratch/exfoliate/move/rotate the flake. For all functions of the microneedle probe technique, we did not observe clear differences when varying the moving speed of the tip. For soldering using a Field's metal, we place a small piece of the metal on the substrate away from the flake and heat up the sample stage above its melting point. Once the metal is melted, we dip the microneedle probe into the melted lump and drag the wire from it towards the flake. For indium soldering (**Fig. 5**), we first pull out a spike by the microneedle from a melting indium bead, and then align it with the flake. Once the spike sticks on the flake at the desired location, we release the spike by raising the microneedle probe.

Tungsten microneedle probes with tip diameters of 50~200 nm are used (KTS, WG-38-0.05) in this paper, and the price for each probe is around 5 USD. **Fig. S2** shows the scanning electron microscope (SEM) images of two microneedle probe tips before and after use. In general, a probe can be used many times to scratch, rotate, move or exfoliate flakes as long as its tip maintains straight or slightly bent. This can be achieved by closely monitoring the position of the tip while slowly lowering it down because we found that whenever the tip touches the substrate, the tip moves slightly in $xy$ direction. If we stop lowering the tip as soon as we find the horizontal movement, the tip will only be slightly bent (**Fig. S2(a)**) whereas if we lower the tip further, it will be bent more (**Fig. S2(b)**). Even when the tip is bent, we can still use it to roll-up or solder metal contacts to flakes since these operations do not require a very sharp tip. We also found that the bent tips are more useful in rolling up the flakes. It is worth noting that after soldering, the probes can have metal residues remain on tips which are sticky to flakes, so they cannot be used for other operations.

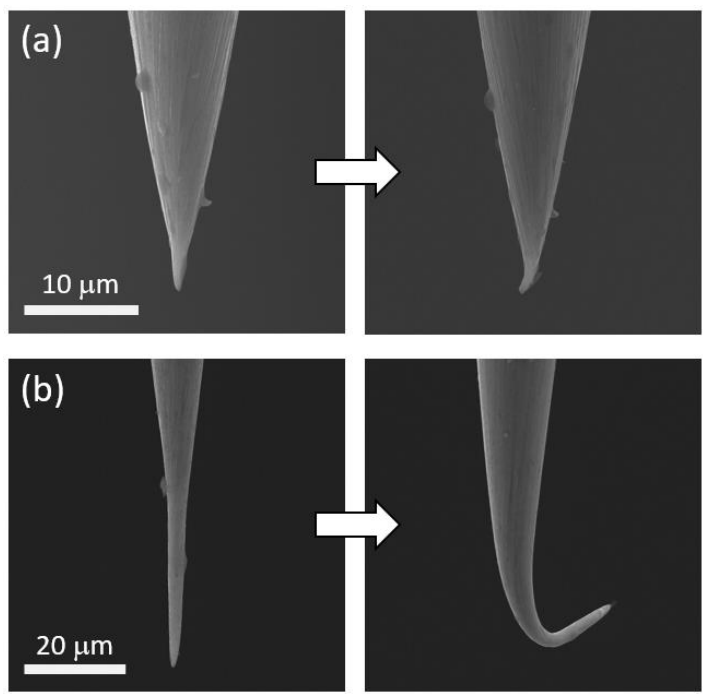

**Figure S2.** SEM images of two new microneedle probe tips that are (a) slightly bent, and (b) bent after use, respectively.

**Electrical measurement**

The 4-terminal resistance of the twisted double bilayer graphene device shown in **Fig. 1(d)** was measured at 1.5 K by applying a small low-frequency (17.777 Hz) AC current of 10 nA between the source and drain terminals and measuring the voltage drop between another two probes using a lock-in amplifier (Stanford Research SR830). All other measurements shown in this study were done at room temperature. In the in-situ electrical measurement shown in **Fig. 2**, the electrical signal of graphene device was measured by using the same lock-in technique described above for **Fig. 1(d)** with AC current of 1 µA. For measuring soldered graphene and $MoS_2$ devices shown in **Fig. 5**, the source-drain DC voltage bias and resulting current are applied and measured by a Keithley 2400 source-meter while the back gate voltage was controlled by another Keithley 2400.